\documentclass[aps,preprintnumbers,nofootinbib,superscriptaddress]{revtex4}

\usepackage{graphicx}
\DeclareGraphicsExtensions{.pdf}

\usepackage{amsfonts,amssymb,amsmath}
\usepackage{amsthm}

\newcommand{\comment}[1]{}
\newcommand{\ket}[1]{\left |  #1 \right\rangle}

\newcommand{\Tr}{{\rm Tr}}
\theoremstyle{plain}

\theoremstyle{definition}

\begin{document}

\title{Unconditionally Secure Bit Commitment with Flying Qudits}

\author{Adrian \surname{Kent}}
\affiliation{Centre for Quantum Information and Foundations, DAMTP, Centre for
  Mathematical Sciences, University of Cambridge, Wilberforce Road,
  Cambridge, CB3 0WA, U.K.}
\affiliation{Perimeter Institute for Theoretical Physics, 31 Caroline Street North, Waterloo, ON N2L 2Y5, Canada.}

\date{\today}

\begin{abstract}
In the task cryptographers call bit commitment, one party encrypts a prediction in a way
that cannot be decrypted until they supply a key, but has only one valid key.   
Bit commitment has many applications, and has been much studied, but
completely and provably secure schemes have remained elusive.
Here we report a new development in physics-based cryptography which gives a 
completely new way of implementing bit commitment that is perfectly
secure.  The technique involves sending a quantum state (for instance one or more photons)
at light speed in one of two or more directions, either along a secure
channel or by quantum teleportation.   Its security proof relies on the 
no-cloning theorem of quantum theory and the no superluminal signalling principle
of special relativity. 
\end{abstract}

\maketitle

\section{Summary}
We report a new form of cryptography that relies on sending a quantum state at light speed, and 
show that it solves a longstanding cryptographic problem.   

\section{Introduction}
\label{sec:intro}

Alice and Bob participate in a stock market that trades at a specific physical site.   
Alice has a way of generating market predictions, which she wishes to demonstrate
to Bob, in such a way that he can verify the accuracy of her predictions post hoc, but cannot 
possibly exploit them before the predicted events occur.   Bob needs a guarantee that 
Alice's predictions were genuinely made at or before the point she claims they were, and
were not retroactively altered post hoc.   
They can solve their dilemma with a suitably secure protocol for {\it bit commitment}.

Much attention has been devoted to the problem of bit commitment,
which is a basic cryptographic task that is important per se, 
and has many applications to other more complex tasks.  
It also has intriguingly deep connections to fundamental physics, which have been
uncovered in the search for bit commitment schemes whose security is guaranteed
by the laws of physics alone (i.e. without any need for extra computational or technological
assumptions).   

Initially, work in this area focussed entirely on protocols based on
non-relativistic quantum mechanics.     
Bennett and Brassard invented the first 
quantum bit commitment protocol \cite{BBeightyfour}, which (as they
noted) is secure against 
both parties given current technology, but insecure if
Alice has a quantum memory.  
Later attempts at unconditionally secure non-relativistic protocols (e.g. \cite{BCJL})  
were ultimately shown to be futile by the celebrated results of 
Mayers \cite{mayersvthree,mayersprl,mayersone} 
and of Lo and Chau \cite{lochauprl,lochau}, later
further elaborated \cite{mkp,darianoetal}, which show
that no unconditionally secure non-relativistic quantum
bit commitment protocol exists.     

This picture changes radically
when we also exploit the signalling constraints implied
by special relativity.   The possibility of 
using relativistic signalling constraints for bit commitment
was briefly discussed by Mayers \cite{mayersprl},  
who suggested that his version of the no-go theorem should also 
apply to relativistic protocols.  One strategy for bit commitment based
on temporary relativistic signalling constraints was indeed shown to
be insecure against quantum attacks \cite{bcms}. 
The discovery \cite{kentrel,kentrelfinite} that relativistic protocols
{\it can} evade the Mayers and Lo-Chau no-go theorems
thus came as a surprise.    
 Encouragingly, such protocols are practical (although
challenging) to implement with existing technology: 
ref.~\cite{kentrelfinite} describes a relativistic bit
commitment protocol in which the parties can maintain 
bit commitments indefinitely, by exchanging classical information at a 
constant -- and presently feasible -- rate between 
two pairs of separated sites under their respective control.   
It is provably secure against all classical attacks and
against Mayers and Lo-Chau's quantum attacks, and is conjectured
to be unconditionally secure.  However, there is as yet no complete security proof
against general attacks.   

Here we introduce a completely new technique for bit commitment, 
which relies essentially on the properties of quantum information in 
relativistic quantum theory -- specifically, on the 
no-summoning theorem \cite{nosummoning}.   
It is provably unconditionally secure. 
In its simplest form it requires only one quantum state transmission
per commitment.  
It illustrates a new way of exploiting cryptographically the 
control over physical information that special relativity and quantum 
theory together allow.   This seems likely to find many
other applications (e.g. \cite{akloc}).   

One interesting potential application of the present work is to
high-frequency financial trading, in which light speed signalling constraints are already
a significant and potentially disruptive 
factor \cite{wgf}.  In a financial context, bit commitment is usually thought of as a
technique for making encrypted predictions that can be decrypted and
checked after the event.   However, in an appropriately regulated
framework, it could also be used to commit parties to trades without 
immediately making the trade public, or even without immediately
informing one of the trading parties (who would have to have agreed to
a regulatory framework in which the risks and disadvantages of such blind transactions 
were appropriately managed and compensated).  The techniques described here
thus illustrate, inter alia, that light speed signalling constraints 
also offer a tool for either traders
or regulators to reshape market dynamics via secure commitments.  In a possible future
world in which conventional cryptosystems are rendered suspect by 
quantum computers or other developments, they may be the only reliably
secure techniques.  

\section{Bit Commitment with Flying Qudits}

We begin by idealizing to simplify the presentation of the
essential idea.   We will relax these unphysical idealized assumptions later.
We suppose that space-time is 
Minkowski and that nature is described by some appropriate relativistic version 
of quantum theory.   
We suppose that both parties, the committer (Alice) and the recipient (Bob),
have arbitrarily efficient technology, limited
only by physical principles.
We suppose that all their operations and communications
are error-free and that Alice can carry out quantum operations
instantaneously.  
We also suppose they agree in advance on some
space-time point $P$, to which they have independent secure access,
where the commitment will commence. 

We suppose they are independently and securely able to access 
every point\footnote{
Alice and Bob should be thought of here as agencies -- which in the
ideal case are represented everywhere in the future light cone of $P$ -- rather than individuals.}   
 in the causal future of $P$, and instantaneously process and
exchange information there, and that each is able to keep
information everywhere secure from the other unless and until they
choose to disclose it.\footnote{
In a more realistic model, which deviates from our idealized scenario
but can still illustrate all the key features of our discussion,
Alice and Bob could both be large independent collaborative groups of people,
with each group having its own independent secure network of quantum
devices and channels distributed throughout a region in such a way
that the two networks do not overlap but both includes sites close to
any point where quantum states may be exchanged.} 
We suppose also that they can securely exchange quantum states 
at $P$, and at other relevant points to which they both have 
secure access.   

We suppose too that Bob can keep a state private
somewhere in the past of $P$ and arrange to transfer it to Alice
at $P$.  Alice's operations on the state can then 
be kept private unless and until she chooses to return information
to Bob at some point(s) in the future of $P$.        
We also suppose that Alice can send any relevant states
at light speed in prescribed directions along secure quantum
channels.   

They also agree on a fixed inertial reference 
frame, and two opposite spatial directions within that frame.    
We  initially simplify further by working in one space and one time
dimension; we set $c=1$ and take $P$ to be the origin in the fixed frame
coordinates $(x,t)$ and the two spatial directions to be defined
by the vectors $v_0 = (-1 , 0 )$ and $v_1 = (1,0 )$.   

Before the commitment, Bob generates a random pure qudit state $\rho \in {\cal C}^d$, chosen
from the uniform distribution, encoded in a physical system which (idealizing again) we take to be pointlike.  
He keeps it private until $P$, where he gives it to 
Alice.    To commit to the bit $i \in \{0,1 \}$, Alice sends the state $\rho$ along a 
secure channel at light speed in the direction $v_i$, i.e. along the line
$L_0 = \{ (-t, t) , t > 0 \}$ (for $0$) or the line $L_1 = \{ (t, t) , t > 0 \}$  (for $1$).

In the simplest implementation
of this protocol, Alice's secure channel may be physically secured -- 
for instance, a shielded region
of free space.   In this case,
to fit the standard model for mistrustful cryptography, we consider
the relevant channels as lying within Alice's secure laboratory. 
Alternatively, if Alice knows in advance the points at which she
wishes to unveil her commitment, she can predistribute entangled
states
between $P$ and these points, and implement a  secure channel by
teleporting the unknown state to a point on the relevant light ray,
broadcasting the classical teleportation signal from $P$. 
Security here is guaranteed since the classical teleportation signals
carry no information about either the transmitted state or (more
importantly) the direction in which it is teleported. 

For simplicity, we consider here the simplest implementation in which
the state is directly securely transmitted. 
Alice can then unveil her commitment at any point along the transmitted light ray.
To unveil a $0$, Alice returns $\rho$ to Bob at some point $Q_0$ on $L_0$;
to  unveil a $1$, Alice returns $\rho$ to Bob at some point $Q_1$ on $L_1$. 
Bob then carries out the appropriate projective measurement to verify that the returned 
qudit is $\rho$; if he gets the correct answer, he accepts the commitment as honestly
unveiled; if not (given that at this stage of the discussion we make
the idealized assumption of no errors), 
he has detected Alice cheating.

\subsection{Security against Bob}

Given our assumptions, the protocol is evidently secure against Bob, who learns nothing
about Alice's choice until the unveiling.   

\subsection{Security against Alice}

\subsubsection{No perfect cheating strategy}

Once Alice has carried out quantum operations of her choice at $P$, 
her strategy for optimizing the probability of successfully unveiling $0$ is independent of
the unveiling point $Q_0$ on $L_0$ (and similarly for $1$).   Although Alice is free to 
choose the points $Q_0$ and $Q_1$, and may vary them depending on other relevant
information that reaches her from the relevant past light cones, we
may thus without loss of generality consider $Q_0$ and $Q_1$ as
fixed.  

Whatever operations and strategies she chooses, Alice cannot guarantee both that she 
will be able to unveil successfully at $Q_0$ if she (at $Q_0$) chooses to, and also that
she will be able to unveil successfully at $Q_1$ if she (at $Q_1$) chooses.   If she could, she 
would be able to guarantee a successful unveiling at both points, by following the appropriate
strategies at both points.   This would 
violate the no-cloning theorem, since she would be able to guarantee producing
two copies of the unknown pure state $\rho$ in the frame in which $Q_0$ and
$Q_1$ are contemporaneous.   More fundamentally, from an intrinsically relativistic perspective,
she would be violating the {\it no-summoning theorem}
\cite{nosummoning}, an intrinsic feature of relativistic quantum
theory that extends the no-cloning theorem \cite{wz,dieks} and the no-signalling
principle.  
To see this, note that she would be able to guarantee successfully
responding to a  summons from another party at $Q_0$, 
requiring her to produce the state $\rho$ at $Q_0$, and also to a similar summons at $Q_1$ --
and hence to both summonses, which is impossible. 

\subsubsection{Full security against cheating}

Suppose Alice decides, at (or in the past of) $P$, that she wishes to retain as much freedom
as possible in her choice of which bit to unveil, and is willing to accept that this may entail
some risk of being caught cheating.   Specifically, she wants to design a strategy which gives
her a probability $p_0$ of successfully unveiling the bit $0$ at $Q_0$, should she (at $Q_0$)
decide she wishes to, and a probability $p_1$ of successfully unveiling the bit $1$ at $Q_1$, should she (at $Q_1$)
decide she wishes to.    

Any quantum bit commitment protocol in which an honest committer can be certain of successful
unveiling allows such strategies, in which the value of the unveiled bit is genuinely undetermined
(not merely unknown to Alice) until the unveiling, for any probabilities $p_0$ and $p_1$
with $p_0 + p_1 = 1$.    To achieve this, the committer simply has to prepare a state of 
the form 
$$
\sqrt{p_0} \ket{0}_A \ket{0}_I  + \sqrt{p_1} \ket{1}_A \ket{1}_I  \, ,
$$
where the $\ket{}_I$ state is input as the committed bit and the entangled 
$\ket{}_A$ state is stored until unveiling and then measured in the computational
basis.\footnote{See e.g. Refs. \cite{kentbccc,kentshort}.}    Indeed, even classical bit commitment protocols can achieve a 
similiar outcome if the committer chooses their input randomly, without observing the 
random choice.   In this case there is a definite committed bit value (and this is an
important difference in some applications), but from Alice's perspective the unveiled
bit remains unknown, with subjective probabilities obeying $p_0 + p_1 = 1$. 

A sensible definition of cheating thus requires that a cheating Alice can at
least ensure that $p_0 + p_1 > 1$.   
More precisely, we say a protocol with a security parameter $N$ is unconditionally secure provided that,
for any committing and unveiling strategies, we have $p_0 + p_1 < 1 + \epsilon (N)$,
where $\epsilon(N) \rightarrow 0$ as $N \rightarrow \infty$.  

In our protocol, the security parameter is the qudit dimension $d$.   Again, we can treat $Q_0$
and $Q_1$ as fixed.   Suppose that Alice carries out some fixed operations at $P$, with a 
view to giving herself some chance of successful unveiling at either $Q_0$ or at $Q_1$. 
Let $\rho_0$ be the state she generates at $Q_0$ if she subsequently follows the strategy that
optimizes her chances of successfully unveiling there, and $\rho_1$ the corresponding state
at $Q_1$.    Again, we can treat Alice's decisions whether to 
unveil at $Q_0$ and $Q_1$ as independent, and imagine that she chooses
to unveil at both points.    She then produces, at space-like
separated points, the states $\rho_0$ and $\rho_1$ (which may and
generally will be mixed), which, roughly speaking, are intended to 
be near-copies of the initial pure state $\rho$ that are as faithful
as possible.   More precisely, since Bob tests the returned states 
$\rho_i$ by measuring the projector $P_{\rho}$ onto the pure state $\rho$, the probability $p_i$ of each state
$\rho_i$ passing Bob's test is $\Tr ( \rho \rho_i )$.   
We thus have 
$$
p_0 + p_1 = \Tr ( \rho \rho_0 ) + \Tr ( \rho \rho_1 ) \, . $$
Effectively, in a reference frame in which $Q_0$ and $Q_1$ have the
same time coordinate, Alice is attempting to implement $ 1 \rightarrow 2$
approximate cloning of the unknown state $\rho$.  The 
expression on the right hand side, $\Tr ( \rho \rho_0 ) + \Tr ( \rho \rho_1 )$, is a standard measure of
fidelity  for this task.   If Alice's process ensures that $\Tr ( \rho \rho_0 ) = \Tr ( \rho
\rho_1 )$, she is implementing symmetric $1 \rightarrow 2$ cloning;
otherwise, her scheme is asymmetric.   Optimality bounds on the
fidelity have been proved\cite{gm,bem,keylwerner,werner,bbh,cerf,ffc,iblisdiretal,iag}
in both cases.  The symmetric bound suffices, since
any asymmetric scheme can be symmetrized by randomization, without altering
  $p_0 + p_1$ (or in the $1 \rightarrow N$ case, without altering $\sum_{i=0}^{N-1} p_i$) , so there must be a symmetric scheme that is optimal
by this measure.  However, in 
the $1 \rightarrow 2$ case it seems a little more satisfying to use
the explicit form for the asymmetric bounds. 
For $1 \rightarrow 2$ qudit cloning they give
$$p_0 + p_1 \leq 2 - \frac{d-1}{d} (a^2 + b^2 ) \, 
$$
where $a^2 + b^2 + \frac{2ab}{d} = 1$.
Optimising via a Lagrange multiplier we find  
$$ p_0 + p_1 = \Tr ( \rho \rho_0 ) + \Tr ( \rho \rho_1) \leq  1 + \frac{2}{d+1} \, . $$
We thus indeed have unconditional security: the expression $( p_0 + p_1 - 1 )$ (Alice's ``wiggle room'') 
is bounded by a term that is $O( d^{-1})$, i.e. $O ( \exp ( - Cn ))$, where $n$
is the number of qubits that Alice and Bob exchange.  

\subsection{The non-ideal case} 

Realistically, the qudits
used will not be pointlike physical systems, it will not be 
possible for Alice and Bob to have secure access to precisely the
same space-time point to exchange a qudit, and 
Alice may not be able to transmit qudits at precisely
speed $c$. \footnote{Even if they are encoded by photons, they may
be sent through some medium with lower light speed.}  
Realistically, too, Alice's and Bob's operations will take nonzero time. 
All of these limitations effectively
introduce time delays at various stages of the protocol.

These delays can all be allowed for in a realistic definition of the task\footnote{Cf.~the
corresponding discussion for the no-summoning
theorem.\cite{nosummoning}}, without making the 
protocols insecure.  However, they do affect the interval over which the
commitment is guaranteed to be secure.   
To see this, suppose that Alice and Bob
attempt to implement a protocol which ideally prescribes that
the qudit should be returned at one of the points $Q_0 , \ldots ,
Q_{n-1}$.  Let $Q'_0 , \ldots , Q'_{n-1}$
be the points at which Bob would actually be 
able to complete his tests, given a real (non-ideal) implementation
of the protocol, which allows Alice to return the qudit 
in specified finite regions of space-time to the future of the points
$Q_i$.    Bob can then only be confident that Alice was committed
at points lying outside the past light-cones of the $Q'_i$ (i.e. that she
has no guaranteed successful strategy for changing her commitment 
that relies on operations at any such points).   
In other words, although Alice may need in fact to be committed at
the point $P$ (and hence all points in its causal future) in order 
to implement her side of the protocol, the security analysis
will only persuade Bob 
that she was committed by some point (or in some region) that lies in the 
future of $P$.\footnote{As with other relativistic protocols, the fact that the local space-time geometry
in the vicinity of the Earth is not exactly Minkowski can also be
allowed for.\cite{kentrelfinite}}

If the $Q'_i$ are suitably close to the $Q'_i$, this need not make any substantial
difference, and thus may not be particularly relevant unless great precision over the place and time at which Alice 
became committed is required. However, when -- as may be the case in our stock market
example -- a relatively precise timing of Alice's commitment {\it is} potentially
crucial, it will be desirable to try to minimize the time delays arising
in practical implementations.  

\subsection{Channel losses and errors}

In reality, Alice and Bob's operations will not be error-free.   
With some nonzero probability, the qudit state will be 
altered somewhere between Bob's sending it to Alice and Alice
returning it to Bob.  Also, with some nonzero probability, the
qudit will either be lost somewhere between Bob's transmitting it at some
point $P'$ in the (presumably) near past of $P$ and his receiving
it at $Q'_0$ or $Q'_1$ (because Bob's or Alice's channels are not
perfect) or will fail to be detected when received
at $Q'_0$ or $Q'_1$ (because Bob's detectors are not perfect).  

The possibility of losses can be countered by running $N>1$ copies of
the protocol, timed so that they are effectively run in parallel.
Instead of supplying a single random qudit to Alice at $P$, Bob
supplies a labelled sequence of $N$ independently random qudits
$\rho^{1} , \ldots , \rho^N$, within a time interval short 
compared to the times and distances separating $P$ and $Q'_i$ (in 
lab frame).   To commit the bit value $i$, Alice is required to 
send all the qudits to the point $Q_i$, and return them from there,
with their original labels, to Bob at $Q'_i$.   
Bob accepts the commitment as valid so long 
as his tests identify $M$ qudits as those he originally sent,
where $M$ is statistically significantly above $N/2$, in the 
sense that the probabilities $p'_0$ and $p'_1$ for Alice being
able to satisfy this test at $Q'_0$ and $Q'_1$ satisfy 
$ p'_0 + p'_1 \leq 1 + \epsilon(d,N,M)$, where 
$\epsilon(d,N,M)$ is suitably small (and can be made arbitrarily small by suitable
parameter choices).    

This strategy of redundant parallel commitment also gives a way of 
countering channel and detector errors.   Standard (and more
efficient) error
correction techniques can also be used for this purpose.   

The possibility of losses and errors is always an issue in
quantum cryptography, and always adds
to the practical challenges in implementing protocols.
The protocols considered here are, of course, no exceptions.  
However, errors and losses raise no qualitatively new
problems {\it of principle} for the protocols considered here.   
It is not the case, for example, that any nonzero error or 
loss rate makes it impossible to implement a secure version
of the protocol.  On the contrary, provided
the loss and error rates are not too large, they can 
be countered by standard error correction
techniques, reducing the probability of false positives (Alice
successfully cheating) and false negatives (Bob being unable to 
accept Alice's honest unveiling) to arbitrarily small agreed levels --
which is the best scenario possible in practical cryptography.   

\subsection{Relation to standard mistrustful cryptography}

Note that relaxing our idealized assumptions allows us to fit the task definition 
into a standard cryptographic model \cite{kentrelfinite}  for mistrustful 
parties in Minkowski space-time.   The non-idealized case allows us to
drop the assumption that Alice and Bob each have independent secure
access to every space-time point.   Instead, we can adopt the standard
assumption that Alice and Bob
control suitably configured disjoint regions of space-time, their ``laboratories''. 
Each trusts the security of their laboratory and all devices contained
within it, but need not trust anything outside their laboratory.   

For example, we can take Alice's laboratory to be a  
connected region of space-time that includes, near its boundary, $P$ and all allowed
unveiling points $Q_i$, and includes light ray segments corresponding
to the secure channels joining $P$ to each
$Q_i$: this allows Alice to receive a state at $P$ and transmit it
securely to any $Q_i$.    We can take Bob's laboratory to be a disjoint connected region
of space-time that includes a point $P'$ in the near causal past of
$P$, from which he sends the unknown state to $P$, and points $Q'_i$ in
the near causal future of each possible unveiling point $Q_i$, to
which Alice is 
supposed to send the unveiled state if she unveils at $Q_i$. 
This allows Bob to generate the unknown state securely, transmit it
to $P$, and then Alice to transmit it securely to some $Q_i$ of her choice, 
and return it to Bob at $Q'_i$, where he can test it securely.  (See Figure 1.)

As is standard in mistrustful cryptographic scenarios, we
assume that Alice and Bob are the only relevant parties -- no one else
is trying to interfere with their communications -- and that they 
have classical and quantum channels (which in principle can be made
arbitrarily close to error-free) allowing them
to send classical and quantum signals between the relevant points.   

It is perhaps worth stressing here that, as with all such models, the main
purpose is to show that the task is well-defined and implementable
according to standard definitions.  The model is not meant to
prescribe how the protocol must be implemented in practice: for example, if 
Alice uses teleportation to transmit the qudit, her labs need not be connected.
The point here is only to note one sensible way of implementing the protocol that uses only standard
cryptographic assumptions.

\begin{figure}[t]
\centering
\includegraphics[scale=0.5]{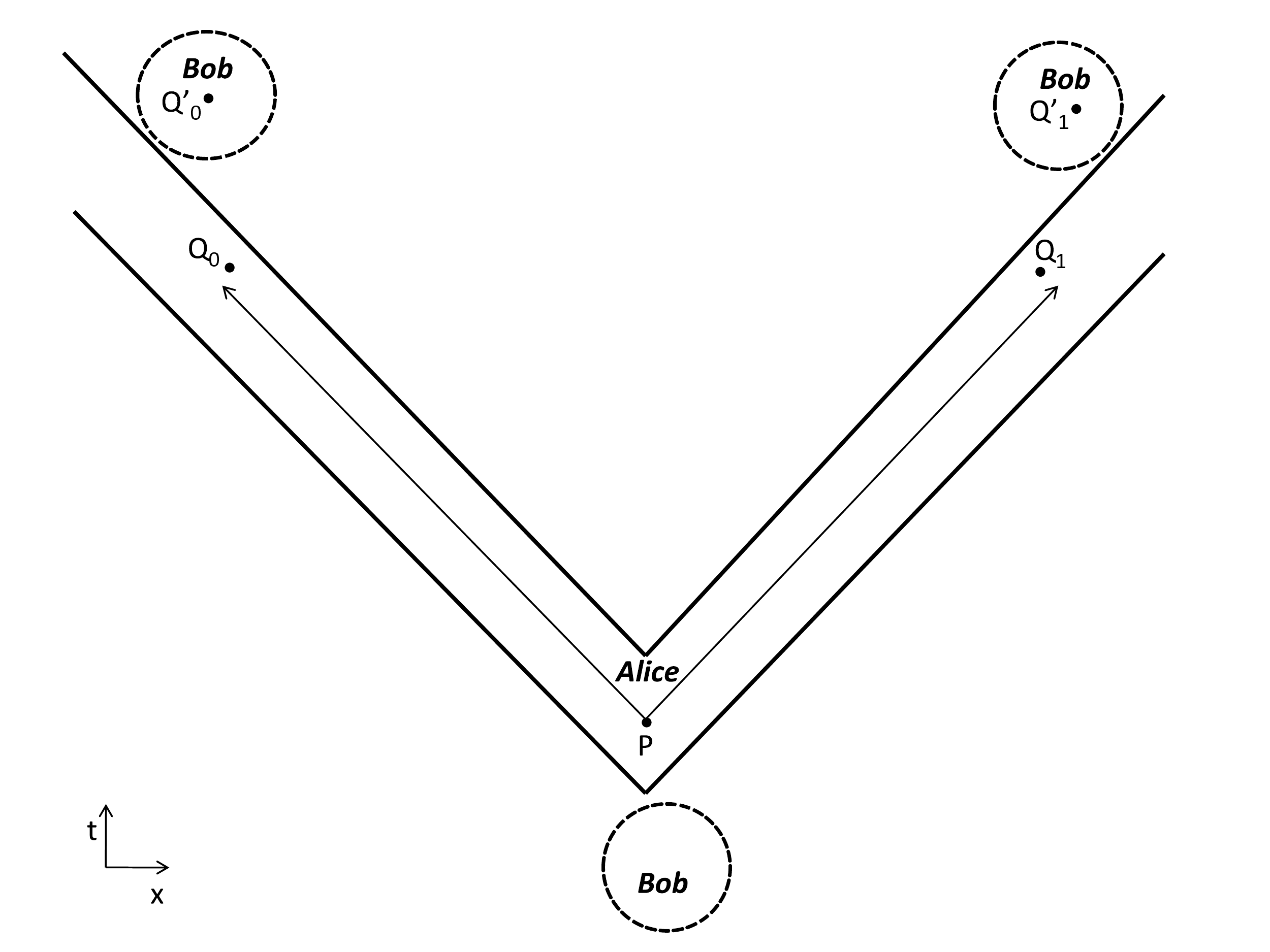}
\caption{One implementation of the non-ideal protocol in 
$1+1$
dimensions.   (Not to scale.)  Alice and Bob control disjoint regions of space-time,
representing their respective secure laboratories.   Bob sends the
unknown state to Alice at point $P$, and she propagates it at near
light speed to one of 
the points $Q_i$, from where she sends it for Bob to receive at
$Q'_i$.   The arrows within Alice's lab denote light rays; the separations from $P$ to the $Q_i$ are
timelike to allow for delays at either end as well as possible subluminal 
communication.   Note that, while Bob's laboratories here are disconnected, in
more than one spatial dimension they can be connected without
overlapping with Alice's. } 
\end{figure}

\subsection{Three dimensions and Committing more Data} 

We now consider three space dimensions, and suppose that Alice wants to commit
to a value in $\{ 0 , \ldots , m -1 \}$, for some integer $m \geq 2$.
To achieve
this we generalize the above protocol.  We fix $m$ distinct lightlike directions from the 
commitment point $P$, defined by $m$ distinct vectors $v_i$ in 
the agreed fixed inertial frame.\footnote{For efficiency, if no other
relevant constraints apply, the vectors should be roughly equally
separated.}     To commit to the value $i$, Alice sends the 
received qudit from point $P$ at light speed in the direction $v_i$.   To unveil
her commitment, she returns the qudit to Bob at some point $Q_i$ on the 
relevant forward light ray from $P$.    By the reverse triangle
inequality for future-oriented timelike and lightlike vectors in
Minkowski space, 
any such point $Q_i$ is spacelike separated from any other such $Q_j$ (for $j \neq i$), 
and so Alice's decisions about unveiling at $Q_i$ cannot be influenced by her 
communicating to $Q_i$ information she learns at any point $Q_j$ on
any of the other light rays.   

Suppose that Alice carries out some fixed operations at $P$, with a 
view to giving herself some chance of successful unveiling at each of 
the points $Q_i$.  Again, without
loss of generality, we can analyse security while treating the $Q_i$ as fixed
and (since an unveiling at $Q_i$ can be postponed to
any future point on the relevant light ray) assuming they have
the same time coordinate in some inertial frame $F$.   
Let $\rho_i$ be the state Alice generates at $Q_i$ if she subsequently follows the strategy that
optimizes her chances of successfully unveiling there.    
Again, we can treat Alice's decisions whether to 
unveil at $Q_i$ as independent of her actions at $Q_j$ for $j \neq i$, and imagine that she chooses
to unveil at all $m$ points.   In the frame $F$, her actions then
become an attempt at $1 \rightarrow m$ cloning.  

The optimality bound \cite{gm,bem,keylwerner,werner}
for symmetric $1 \rightarrow m$ qudit cloning implies, 
via the symmetrization argument mentioned earlier, that 
$$ \sum_i p_i =  \sum_i \Tr ( \rho \rho_i ) \leq  1 + O( 2 m d^{-1} )   \, . $$
We thus have unconditional security in this case also: the expression $( \sum_i
p_i - 1 )$ 
is again bounded by a term that (for fixed $m$) is $O( d^{-1})$, i.e. exponentially
small in the number of qubits used.      

In principle, this strategy of direction-dependent commitment allows
an arbitrary amount of data to be securely committed and unveiled, using a single qudit, 
and within a fixed finite space-time volume.  
However, there are two important caveats here. 

First, security requires that $2 m d^{-1}$ is small, meaning that 
the dimension $d$ needs to scale at least\footnote{One might choose
tighter security criteria in this case.} linearly in $m$. 
Second, Alice needs to be able to transmit from $P$ in $m$ distinct
directions, and Bob needs to be able to distinguish $m$ distinct 
unveiling points on the relevant light rays.  Alice thus needs to
be able to specify transmission directions to within a solid angle\footnote{Or angle,
if the relevant directions lie in a plane.} small compared to $m^{-1}$, and Bob needs to be
able to distinguish separations small compared to the distances 
between the corresponding unveiling points.   These precisions
are thus exponential in the number of bits committed, $\log_2 m$.   In realistic
implementations, attaining such precisions should be considered as consuming resources
that also scale at least exponentially in $\log_2 m$ for large $m$.  

\section{Discussion}

The new bit commitment protocols described here are theoretically
interesting in that they make use of a property of relativistic
quantum theory -- which seems most fundamentally characterised
by the no-summoning principle \cite{nosummoning} -- that has not previously been exploited
cryptographically, and which looks likely to find many other
applications.   Working out in full generality precisely which applications of 
bit commitment can be implemented using sequences of these protocols 
poses some interesting theoretical and practical challenges.   

To implement the protocol efficiently and reliably requires
transmitting qudits efficiently and reliably (by teleportation,
or within cryptographically secure regions of space), at near light speed
along pre-determined alternative paths.   
If one takes an optimistic view, common among quantum information scientists, 
technological barriers that currently prevent near-ideal
implementation of quantum information processing and communication
will ultimately be overcome.  At some point, the transition to 
practical quantum information processing and communication
will take place -- and at that point the practicality of quantum computing
motivates a transition from (newly vulnerable) classical cryptographic
protocols to secure quantum protocols.   This is sufficient
motivation for the protocols described here: if and when practical and
reliable quantum information processing technology emerges, they will
be both practical and practically relevant.    

That said, it is worth reviewing how practical the protocols already are
with present technology.  For a proof that the basic concept can be implemented, Alice 
needs only be able to route photons at near light speed along 
alternative paths -- which may be through free space -- in regions
which we can assume lie within her laboratory (and thus are secure), 
and return the photons to Bob's detectors.   Moreover, for a partial proof
of concept, one might initially accept an unreliable implementation,
in which Alice's commitments are sometimes verified by Bob but 
sometimes (because of losses or errors) unverified.     
One might too accept implementations in which the photons travel at 
significantly lower than light speed (along optical fibres of high
refractive index): such implementations still guarantee a finite
(though shorter) duration commitment, so long as the allowed unveiling points
are space-like separated.   
Under some or all of these relaxed assumptions, the protocol seems 
well within the scope of current technology.   How reliably
it can presently be implemented, over what ranges, with what
levels of security, and how small the transmission and other delays
can be made compared to the ideal protocol, are open questions, 
which we offer as challenges to the ingenuity of experimentalist 
colleagues.     

Returning to the optimistic long term view of quantum information
technology, it seems to us that a plausible future cryptographic environment
will require unconditional security for bit commitments
of short duration, and that these protocols may turn out to be the most efficient and easiest
solutions.   For example, one can imagine short term commitments
relating to a stock market being made and unveiled by two parties
exchanging photon signals that are transmitted along 
independently secure links.   
For such applications, it will be particularly interesting to explore and
analyse the security and efficiency of 
chaining protocols and redundant encoding (discussed in the
Appendix). 

Previous protocols \cite{kentrel,kentrelfinite} that make use of the
Minkowski causal structure to implement secure bit commitment rely on the
fact that data introduced at one site is not available to
an agent at a space-like separated site.   As noted earlier, these protocols
are immune to Mayers' and Lo-Chau's cheating strategies. 
The MLC strategies do, in principle, define an operation
that the committer could carry out in order to cheat the protocols of
Refs. \cite{kentrel,kentrelfinite} by altering their
commitment. However, this operation always depends on data introduced
at a space-like separated site, and so is never knowable by the 
committer.  

The protocols introduced here highlight another significant limitation 
of Mayers' and Lo-Chau's no-go theorems that appears to have
previously gone unnoticed.   Namely, 
in relativistic quantum theory, the unitary operation required for a MLC attack can be
known to both parties but impossible to implement physically, as it
represents a spacelike translation that would violate causality.  
This reinforces the point -- if any reinforcement
were needed -- that these celebrated and fundamental 
theorems are correctly understood as applying specifically to protocols
that rely only on non-relativistic quantum mechanics.    
They can be extended to some important classes of relativistic protocols -- for
example those in which the unveiling point takes place at a fixed
point in the causal future of all the operations carried out in the
commitment phase -- but do not apply to general protocols based on
relativistic quantum theory.  

Like all intrinsically quantum bit commitment protocols \cite{kentbccc,kentshort}, the protocols
considered here allow the committer to commit a bit in quantum
superposition, which is ``collapsed'' to a classical bit value only
when unveiled.   For some important applications of bit commitment,
this feature makes no essential difference.  For example, if the bit
commitment encodes a prediction, allowing the committer to make a 
superposition of predictions merely gives them the freedom to add
a random element to their prediction -- which is also possible with
any classical bit commitment scheme, since the committer can always
randomize their input.   Indeed, for some intrinsically quantum applications, 
in which the ultimate aim is to unveil a quantum state, 
it may be a positive advantage that the scheme allows commitment of 
a qubit rather than a bit.   On the other hand, our scheme cannot be
securely used in applications of
bit commitment in which it is crucial that the committed bit takes
a classical value fixed at (or before) commitment.   
It bears reemphasizing that this is a quite general feature of
intrinsically quantum bit commitment schemes, and quite separate from
the cheating possibilities pointed out by Mayers and
Lo-Chau.\footnote{If (which we disrecommend) one were to break with established convention 
and define quantum bit commitment so as
to require a guarantee of the classicality of the committed bit, it 
would become almost trivial to show that quantum bit commitment is
impossible \cite{kentbccc}, and Mayers' and Lo-Chau's impossibility
proofs for non-relativistic quantum bit commitment would be
unnecessary.}

The protocols we have described are unconditionally secure, but not
necessarily optimally efficient, in the sense that they achieve
optimal security for given resources.   It will be interesting to investigate the range of 
possible strategies and their efficiencies.   One particularly
interesting possibility is to consider protocols based on summoning
an entangled state.   For example, Bob could generate a randomly
chosen maximally entangled pair of qudits, and initiate the commitment
protocol by giving one of them to Alice -- who must later return it 
as above -- while he keeps the other.   To verify the unveiled 
commitment in this case, he needs to recombine his stored qudit
with the returned qudit at a single site using secure quantum channels, or else carry out a 
non-local measurement.   Either way, he cannot generally verify with
certainty at the point of unveiling, but can do so at a future
point.\footnote{One might hope by this method to obtain a security bound
of $O( d^{-2} )$ instead of the $O( d^{-1} )$ bound obtained above;
however we are not aware of any analysis to date, so this may be an
interesting open question.  Implementations using non-maximally entangled 
states also seem worth analysing.} 

\acknowledgments
I thank Adam Brown for drawing my attention to Ref. \cite{wgf} and
Charles Bennett, Serge Massar, Stefano Pironio and Jonathan Silman for
very helpful comments. 
This work was partially supported by a Leverhulme Research Fellowship,
an FQXi mini-grant, a grant
from the John Templeton Foundation, and by Perimeter Institute for Theoretical
Physics. Research at Perimeter Institute is supported by the Government of Canada through Industry Canada and
by the Province of Ontario through the Ministry of Research and Innovation.


\section{Appendix: Chaining commitments} 

\subsection{Determining viable unveiling locations: some
  scenarios} 

These protocols have the advantage of simplicity: they
require only acting on a qudit with a simple quantum
operation to determine its transmission direction, and then
transmitting it.  The fact that the unveiling location needs
to be on a light ray from the commitment point, in a direction
depending on the committed bit, is, however, potentially a 
disadvantage.   Whether it is problematic, and if so how much so, very much depends on the scenario in which
the bit commitment is being used.   

For example, if the trading
centre of a stock market is physically localized around the 
point $P$, and Alice wants to commit at time $0$ predictions of
prices at a pre-agreed time $t>0$, which she is happy to unveil as soon as
the data cannot be exploited by Bob, the protocol may be perfectly
adequate, so long as they can securely exchange information at 
suitable points distance $(t/2)$ from P in stock market rest frame
(i.e. so long as $t$ is not too large).   

On the other hand, if Alice makes some prediction at time $0$ which
she may or may not wish to unveil, depending on details of the 
market's behaviour after $t=0$ -- perhaps, for example, because the prediction
is of some future events conditioned on others, and she wishes to
give away no information about her predictive abilities unless
the conditioning events take place -- then these protocols alone
do not suffice.   Alice's decision to unveil on any light ray from $P$
must necessarily be made in ignorance of events at the location of $P$ at later
times.   

\subsection{Chaining commitments}

Some flexibility in the location of the unveiling point, relative to
the commitment point, can be achieved by chaining a number of 
our commitment protocols in sequence.   For example, consider the
following extended protocol.    Alice and Bob fix $P$, as above, 
and a time interval $T$ in their fixed inertial frame; they
also fix the qudit dimension $d$.  At $P$, Bob initiates the
protocol by giving Alice an unknown qudit, and she commits 
a bit by sending it along the appropriate light ray to 
points $Q_0$ or $Q_1$, which have time coordinate $T$. 
Suppose, for example, she commits to $0$.  At $Q_0$, she
returns the qudit to Bob, after acting on it with a 
randomly chosen $d$-dimensional teleportation operation.
This operation is specified by an integer $j$ in 
$\{ 0 , \ldots , d^2 -1 \}$.   Bob initiates a new data
commitment protocol at $Q_0$, giving an Alice a 
qudit of pre-agreed dimension $d'$ (here $d' > d$), 
which is used to commit to the value $j$.   
At $Q_1$, Alice and Bob go through the same operations,
but here Alice simply returns a randomly chosen dummy
qudit to Bob, and commits to a randomly chosen value $j'$. 
This procedure can be iterated any number of times.  

To unveil, Alice returns the qudits from the final commitment
without any randomization operation, allowing Bob to infer
the operations necessary to derandomize the qudits returned
earlier, and hence the originally committed bit.

This procedure has the advantage that Alice's final unveiling
location (although randomly determined) will generally be 
timelike rather than lightlike separated from $P$.   By 
varying the parameters of the chained protocol
the distribution of separations can be optimized (within the
space of possible distributions) for any given task.
Of course, chaining the protocol requires extra quantum
operations and communications from both parties, 
which moreover grow exponentially in the number of
iterations.   It may perhaps be possible to mitigate
this by adapting techniques used  \cite{kentrelfinite} to eliminate exponential blow-ups
for classical relativistic bit commitment protocols \cite{kentrel}, which allow one to 
truncate successive linked commitments after
a fixed number of iterations and replace them by newly started commitments, while 
retaining security.   

This strategy remains to be fully explored.  
We mention chained protocols here to note that the strategy of 
making bit commitments by appropriate light speed 
transmission --- or indeed using similar techniques for other
cryptographic applications --- is not as inflexible
as it may initially seem from considering the 
simplest (unchained) protocols. 
We leave the analyses of resource optimization,
practicality and security for chained protocols for 
a future discussion.

\subsection{Redundant encoding}

Another valuable strategy is to run two or more 
implementations of the protocol in parallel, using different
bit codings.  For example, in the basic one-dimensional
protocol, Bob can give Alice two independently randomly
generated unknown pure states, $\rho_0$ and $\rho_1$, at
the same point $P$.  She follows the protocol described
above with $\rho_0$, sending it along the line $L_i$ to
commit to bit value $i$; she follows the protocol with
reversed conventions for $\rho_1$, sending it along the
line $L_{i+1}$ (where $i+1$ is defined modulo $2$) to
commit to bit value $i$.   
This allows her to unveil the commitment at any point
$Q_0$ on $L_0$ or any point $Q_1$ on $L_1$, regardless
of the committed bit value.   To prevent her cheating, 
she must also be required to unveil at some point on the opposite
line.  

Giving Alice this freedom ensures that the bit can
be unveiled on an agreed light ray -- or indeed, if
the parties wish, at a pre-agreed point on that
light ray.   Note that if Bob accepts such an unveiling 
immediately Alice has scope for a form of 
temporary cheating.\footnote{``Temporary'' in the sense that at any spatial coordinate
in any given inertial frame it will 
be at most temporarily effective. }  For example, she could cheat by sending
both $\rho_0$ and $\rho_1$ along the agreed light
ray $L_0$ and then decide which to return to Bob
at the unveiling point $Q_0$.   However, any such
cheating will not allow her to produce a consistent unveiling 
at $Q_1$.  Her deception will thus eventually become evident to Bob.\footnote{``Eventually'' in the
sense that, by broadcasting the results of his measurements on
the states returned at $Q_0$ and $Q_1$, and comparing the results
when they arrive at the same point, he can ensure that he will be aware of cheating 
at any or all points in the intersection of the causal futures of
$Q_0$ and $Q_1$.}

In many scenarios -- for example when individual commitments are part
of an ongoing series of transactions of value to both parties --
the opportunity to deceive Bob temporarily is of
little value to Alice.  Of course, in scenarios where Bob can suffer significant
loss through temporary deception, and cannot recover it once the
cheating is exposed, he should not accept Alice's unveiling
until (i.e. at spacetime points where) he can combine all the relevant
unveiling data. 

Note that -- with the above caveats about temporary cheating -- 
redundant encoding can be combined with the chaining strategy
described above, allowing Alice and Bob to ensure that the 
original commitment is unveiled {\it at any point they wish} in the 
causal future of $P$.   In our stock market example, for instance,
they can ensure that the commitment is unveiled at the market location
at any agreed future time.

\end{document}